\documentclass[useAMS,usegraphicx]{mn2e}
\usepackage{rotate}
\usepackage{times}
\newif\ifAMStwofonts
\AMStwofontstrue

\def\gs{\mathrel{\hbox{\rlap{\hbox{\lower4pt\hbox{$\sim$}}}\hbox{$>$}}}}
\def\ls{\mathrel{\hbox{\rlap{\hbox{\lower4pt\hbox{$\sim$}}}\hbox{$<$}}}}

\def\chandra{{\it Chandra}}

\def\rosat{{\it ROSAT}}

\def\xmm{{\it XMM-Newton}}

\def\et{{et al.\ }}

\def\pks{{PKS~2004--447}}
\def\3c{{3C~273}}

\def\chidof{{\chi^2_\nu/{\rm dof}}}
\def\ka{{K$\alpha$}}

\def\oiii{{[O~\textsc{iii}]}}
\def\hbeta{{$\rm H\beta$}}
\def\halpha{{$\rm H\alpha$}}

\def\feii{{Fe~\textsc{ii}}}
\def\nh{{N_{\rm H}}}

\def\A{{\rm\thinspace \AA}}
\def\cm{{\rm\thinspace cm}}
\def\erg{{\rm\thinspace erg}}
\def\eV{{\rm\thinspace eV}}

\def\ga{{\rm\thinspace gauss}}
\def\Hz{{\rm\thinspace Hz}}
\def\MHz{{\rm\thinspace MHz}}
\def\pHz{{\rm\thinspace Hz^{-1}}}
\def\K{{\rm\thinspace K}}
\def\keV{{\rm\thinspace keV}}
\def\km{{\rm\thinspace km}}
\def\kpc{{\rm\thinspace kpc}}

\def\m{{\rm\thinspace m}}

\def\pc{{\rm\thinspace pc}}
\def\s{{\rm\thinspace s}}
\def\ks{{\rm\thinspace ks}}
\def\ps{{\rm\thinspace s^{-1}}}
\def\W{{\rm\thinspace W}}

\def\cts{{\rm\thinspace count}}

\def\cps{\hbox{$\cts\s^{-1}\,$}}

\def\ergpscmps{\hbox{$\erg\cm^{-2}\s^{-1}\,$}}
\def\ergpscmpspkeV{\hbox{$\erg\cm^{-2}\s^{-1}\keV^{-1}\,$}}
\def\ergpscmpspa{\hbox{$\erg\s^{-1}\cm^{-2}\A^{-1}\,$}}

\def\ergps{\hbox{$\erg\s^{-1}\,$}}

\def\kmps{\hbox{$\km\ps\,$}}

\def\pscm{\hbox{$\cm^{-2}\,$}}

\def\WpHz{\hbox{$\W\pHz\,$}}

\title[The SED of \pks]
      {
The spectral energy distribution of \pks: a compact steep-spectrum source
and possible radio-loud narrow-line Seyfert 1 galaxy
      }

\author[L. C. Gallo \et]
       {L. C. Gallo,$^{1,2}$ 
	P. G. Edwards,$^2$ 
	E. Ferrero,$^3$ 
	J. Kataoka,$^4$ 
	D. R. Lewis,$^{5,6}$ 
	S. P. Ellingsen,$^6$ 
\newauthor 
	Z. Misanovic,$^1$ 
	W. F. Welsh,$^7$
	M. Whiting,$^5$ 
	Th. Boller,$^1$ 
	W. Brinkmann,$^1$ 
\newauthor 
	J. Greenhill$^6$ 
	and A. Oshlack$^{8,9}$ \\ 
$^{1}$ Max-Planck-Institut f\"ur extraterrestrische Physik, Postfach 1312, 85741 Garching, Germany \\
$^{2}$ Institute of Space and Astronautical Science, Japan Aerospace Exploration Agency, 3-1-1 Yoshinodai, Sagamihara, Kanagawa 229-8510, Japan\\
$^{3}$ Landessternwarte Heidelberg, K\"onigstuhl 12, D-69117, Heidelberg, Germany \\
$^{4}$ Department of Physics, Tokyo Institute of Technology, 2-12-1 Ohokayama, Meguro, Tokyo 152-8551, Japan \\
$^{5}$ Australia Telescope National Facility, CSIRO, PO Box 76, Epping, NSW 1710, Australia \\
$^{6}$ School of Mathematics and Physics, University of Tasmania, Private Bag 37, Hobart, TAS 7001, Australia \\
$^{7}$  Department of Astronomy, San Diego State University, San Diego, CA 92182, USA \\
$^{8}$  School of Physics, University of Melbourne, Parkville VIC 3010, Australia\\
$^{9}$  current address: Bioinformatics Division, Walter and Eliza Hall Institute of Medical Research, Parkville VIC 3050, Australia\\
}
\date{Accepted. Received. }
\pagerange{\pageref{firstpage}--\pageref{lastpage}}
\pubyear{2006}
\begin{document}
\maketitle
\label{firstpage}

\begin{abstract}
An investigation of the spectral energy distribution (SED) of the
compact steep-spectrum (CSS) source and possible radio-loud narrow-line
Seyfert 1 galaxy (NLS1), \pks, is presented.
Five out of six well-studied radio-loud NLS1 share this
dual classification (optically defined as a NLS1 with radio definition of a
CSS or giga-hertz peaked spectrum (GPS) 
source), suggesting that the connection could have a physical origin.
The SED is created from simultaneous observations (within 24 hours) at
radio (from ATCA), optical/NIR (from Siding Spring) and UV/X-ray (from \xmm) 
wavelengths.  The X-ray data show evidence of short-term variability
(primarily a $\sim30$ per cent increase in the final $4\ks$ of the observation),
a possible soft excess, and negligible absorption.  
Together with the rest of the SED, the X-ray emission is excessive in comparison
to synchrotron plus synchrotron self-Compton (SSC) models.  
The SED can be described with a two component model
consisting of extended synchrotron/SSC emission with
Comptonisation in the X-rays,
though SSC models with a very high electron-to-magnetic energy density ratio
cannot be excluded either.
The peak emission in the SED appears to be in the near infrared, which
can be attributed to thermal emission ($T\approx1000\K$) from a dusty torus.  
Analysis of a non-contemporaneous, low-resolution optical spectrum suggests
that the narrow-line region (NLR) is much more reddened than the 
X-ray emitting region
suggesting that the gas-to-dust ratio in \pks\ may be very different than in
our own Galaxy.
This could be achieved if the radio jets in \pks\ deposits material from
the nucleus into the NLR.
Long-term radio monitoring of \pks\ shows a rather constant light curve
over nearly a six month period with the exception of one outburst when
the $6.65$~GHz flux increased by $\sim35$ per cent over 19 days.
It is not possible to differentiate between intrinsic or extrinsic
(i.e., interstellar scintillation) origins for this outburst, but the 
detection of the rare event demonstrates the importance of intensive 
monitoring campaigns.
In comparison to general samples of GPS sources, which appear to be X-ray weak,
NLS1-CSS/GPS sources possess stronger X-ray emission relative to radio (comparable
to normal radio-loud AGN).  In addition, NLS1-CSS/GPS sources also exhibit
lower intrinsic absorption than GPS sources of similar X-ray luminosity.
This is consistent with the additional X-ray component required for \pks, but
larger samples of NLS1-CSS/GPS are needed before any conclusive remarks can
be made.
\end{abstract}

\begin{keywords}
galaxies: active -- 
galaxies: nuclei -- 
quasars: individual: \pks\  -- 
radio continuum: galaxies --
X-ray: galaxies 
\end{keywords}


\section{Introduction}
\label{sect:intro}

Compact steep-spectrum (CSS) and giga-hertz peaked spectrum (GPS) sources
constitute a large fraction of the powerful (log$P_{1.4}\gs25\WpHz$)
radio source population.
Morphologically, they are compact and often only partially resolved on 
milliarcsecond scales, 
with projected
linear sizes of less than $1$ and $20\kpc$ for GPS and CSS, respectively.  
However, these objects are not just radio cores, rather they constitute  
complete systems with small-scale jets and lobes.

Two scenarios have been proposed that attempt to explain the CSS/GPS
phenomenon (see O'Dea 1998 for a review).  The first is that CSS/GPS 
sources form stages of an evolutionary
sequence for all large radio sources.  The more compact objects (GPS) are the
youngest sources which, with time, expand in size to form CSS and eventually 
large radio objects
(the youth scenario). The second picture is
that CSS/GPS sources are kept compact by high density material which
inhibits the radio source from expanding (frustrated scenario).

The radio spectra of GPS sources show curvature with peaks
above $500\MHz$.  CSS objects are characterised by steep spectra
($\alpha>0.5; S\propto\nu^{-\alpha}$) with a low-frequency, often 
undetectable, turnover (e.g. $\nu_{t}\approx100\Hz$).  
Perhaps the strongest evidence linking GPS and CSS sources is the
relationship (anticorrelation) between turnover frequency and
projected linear size (Fanti \et 1990; O'Dea \& Baum 1997).
In fact, it has been suggested that GPS sources evolve into CSS sources,
and perhaps eventually into FR-I/II objects in a self-similar manner
(Snellen \et 2003).

Studies of the class properties of CSS/GPS quasars at other wavebands
are less complete, but to first-order the observations are not
radically different from non-CSS, radio-loud quasars.
The far-infrared $IRAS$ luminosities are consistent with arising from AGN, 
similar to extended, radio-loud quasars (e.g. Heckman \et 1994). 
There is some evidence for an additional component in the near-infrared (NIR),
which could be due to emission from $\sim1000\K$ dust, possibly from
an AGN heated torus (de Vries \et 1998).
Host galaxies are comparable between GPS, CSS, and FR-II 
(e.g. de Vries \et 2000).  Optical spectra are rich in emission lines,
similar to extended radio sources (e.g. Morganti \et 1997) and the
emission is aligned with the radio source suggesting interaction of
the two mediums (e.g. the radio source shocks the emission-line gas)
in the production of lines (e.g. Gelderman \& Whittle 1994; de Vries \et 1998;
Bicknell \et 1997; Labiano \et 2005).
Until recently there has been little focus on X-ray studies of CSS/GPS sources,
primarily because they appear to be X-ray dim (Baker \et 1995).  This could
be because of additional X-ray absorption, which has been detected in many
objects of this class (e.g. Elvis \et 1994; Guainazzi \et 2006; Vink \et 2006).
Recent studies with \chandra\ and \xmm\ show that the X-ray spectra of
GPS source are well characterised by a single power law modified by
intrinsic absorption (Guainazzi \et 2006; Vink \et 2006).
X-ray variability on hourly time scales has only been reported for
one GPS, COINS~J$0029+3456$ (Guainazzi \et 2006).

In this work we examine the spectral energy distribution (SED) of one
CSS source, \pks.
The PKSCAT90 catalog (Wright \& Otrupcek 1990) gives flux density
measurements at frequencies between 408 MHz and 8.4 GHz for \pks.
These were not contemporaneous, but suggest that the slope of the radio
spectrum is $\sim0.1$ at low frequencies, steepening to $>0.5$ at
higher frequencies.

\pks\ also shows evidence for having an angular size of $<1\arcsec$, as
observations on the $6\km$ baseline of the Australia Telescope Compact Array
(ATCA) have revealed that $\sim99$ per cent
of the flux density is unresolved at 8.6 GHz
(Lovell 1997; Tasker 2000). The source is partially resolved on longer
baselines -- on the $275\km$ Parkes--Tidbinbilla Interferometer baseline a
correlated flux density at 2.3 GHz of $0.39$ Jy was measured, in comparison
to a contemporaneous total flux density of $0.68$ Jy. However, the $2.3$ GHz
single-baseline VLBI survey of Preston \et (1985) yielded a correlated
flux density of $0.28\pm0.02$ Jy on a $10,000\km$ baseline, indicating the
presence of a compact, parsec-scale core. The steep spectrum at cm
wavelengths and sub-arcsecond structure match the criteria for CSS
sources (Fanti \et 1990).

A comparison of all reported flux density measurements for the source
indicates that the source displays variability of $\sim20$ per cent on the
timescale of years at cm wavelengths. Moderate slow variability of CSS
sources is not uncommon at these frequencies (e.g., Edwards \& Tingay
2004).

\pks\ has encountered controversy 
(Oshlack \et 2001, hereafter
Osh01; Zhou \et 2003; Sulentic \et 2003) with its optical definition 
as a narrow-line Seyfert 1 galaxy (NLS1).  By optical definition, NLS1
have narrow permitted lines (FWHM(\hbeta)$\le 2000\kmps$), weak \oiii\
($\frac{\oiii}{H\beta}<3$), and strong \feii\ (see Osterbrock \& Pogge 1985 and 
Goodrich 1989).
\pks\ satisfies the first two criteria, but has weak \feii\ (Osh01).
However, there is no formal, qualitative definition for the \feii\ 
strength, which leaves the interpretation open for debate.

It is also important to note that NLS1 are believed to be young
AGN (e.g. Grupe 1996; Mathur 2000).  They possess smaller black hole
masses and higher accretion rates compared to typical Seyfert 1s of similar
luminosities.  They also appear to have super-solar metallicities, and 
many objects have high IR fluxes, which could be attributed to rapid and/or
intense episodes of star formation.

Five other radio-loud NLS1 are known, which have more secure 
confirmation of the NLS1 characteristics: PKS0558--504 (Remillard \et 1986),
RGB J0044+142 (Siebert \et 1999; though the radio-loudness of this object
is questionable, Maccarone \et 2005),
RXJ0134--4258 (Grupe \et 2000), SDSS J094857.3+002225 (Zhou \et 2003;
Doi \et 2006), and SDSS J172206.03+565451.6 (Komossa \et 2006a).
Whalen \et (2006) have recently identified 16 new radio-loud NLS1, but
radio spectra are not yet available. 

For five of the above mentioned objects (including \pks) the 
radio spectra reveal that all but PKS0558--504 are
consistent with the CSS or GPS definition.
This is of interest for three reasons.  Firstly, NLS1 objects as a class
lie on one extreme end of ``eigenvector 1'' -- the anti-correlation between
\oiii\ and \feii\ strength (Boroson \& Green 1992; see also Brandt \& Boller
1997).  NLS1 show the most
extreme \feii\ emission and weakest \oiii, while radio-loud objects
form the other extreme.  Optically selected samples of NLS1 are notoriously
radio-quiet (e.g. Greene \et 2006).  Secondly, NLS1 are considered to be
in the early stages of AGN evolution (Grupe 1996; Mathur 2000), 
as are GPS/CSS sources.  Though based
on very different arguments, the connection between GPS/CSS and NLS1, and
AGN evolution is intriguing.  Finally, while CSS and GPS sources make up
a considerable fraction of radio selected samples ($\sim30$ and $\sim10$ per 
cent,
respectively), the fraction may be significantly higher when considering
radio-loud NLS1.

In this work we scrutinise the CSS/NLS1 source \pks\ across the 
SED using simultaneous X-ray, near-UV, optical,
near-IR, and radio data obtained 2004 April 11--12, with the intention of 
understanding the connection
between CSS and NLS1 sources.  The rest of the paper is organised as follows.
The observations and data reduction are described in Section 2.
Analysis of the radio, and X-ray data are presented in Section 3,
and 4, respectively.  The SED of \pks\ is considered in Section 5.  Discussion
and summary follow in Section 6 and 7, respectively.

\section{Observations and data reduction}
\label{sect:obs}

A value for the Galactic column density toward \pks\ of 
$3.52 \times 10^{20}\pscm$ (Dickey \& Lockman 1990) is adopted in all of the 
spectral fits.  K-corrected luminosities are calculated using a
Hubble constant of $H_0$=$\rm 70\ km\ s^{-1}\ Mpc^{-1}$ and
a standard flat cosmology with $\Omega_{M}$ = 0.3 and $\Omega_\Lambda$ = 0.7.
The redshift of \pks\ is $z=0.24$ (Drinkwater \et 1997) and the
corresponding luminosity distance is 1.2~Gpc.

\subsection{X-ray and UV observations}
\label{sect:xmm}
\pks\ was observed with \xmm\ (Jansen et al. 2001) for approximately $40\ks$
between 2004 April 11-12 (revolution 795; obsid 0200360201).
During this time all on-board instruments functioned normally.
The EPIC pn (Str\"uder et al. 2001) and MOS (MOS1 and MOS2;
Turner \et 2001) cameras were operated in full-frame mode
with the medium filter in place.
The Reflection Grating Spectrometers (RGS1 and RGS2; den Herder et al. 2001)
also collected data during this time, as did the Optical Monitor
(OM; Mason et al. 2001).  Due to the low count rate, the RGS data were 
mostly background dominated and will not be discussed further.

The Observation Data Files (ODFs) were processed to produce calibrated
event lists using the \xmm\ Science Analysis System ({\sc XMM-SAS v6.5.0})
and the most recent calibration files.  
Unwanted hot, dead, or flickering pixels were removed as were events due to
electronic noise.  Event energies were corrected for charge-transfer
losses, and time-dependent EPIC response matrices were generated using the
{\sc SAS} tasks {\sc ARFGEN} and {\sc RMFGEN}.
Light curves were extracted from these event lists to search for periods of
high background flaring, which were subsequently removed.  The total
resulting exposures were $33522$, $40375$ and $40758\s$ for the pn, MOS1 and
MOS2, respectively.

The source plus background photons were extracted from a
circular region with a radius of 35$^{\prime\prime}$, and the background was
selected from an off-source region with a radius of 50$^{\prime\prime}$
and appropriately scaled to the source
region.  Single and double events were selected for the pn
spectra, and single-quadruple events were selected for the MOS.
The spectra were source dominated between $0.3-10\keV$ with a total of
$12239$ and $8675$ source counts collected by the pn and MOS instruments,
respectively. 
Pile-up was negligible in all instruments.
\begin{table}
\begin{center}
\caption{\xmm\ Optical Monitor observations of \pks.  
Filters and corresponding wavelengths 
are given in columns 1 and 2, respectively.  The flux density
in each filter is reported in column 3.
}
\begin{tabular}{ccc}
\hline
(1) & (2) & (3) \\
OM Filter & $\lambda$ &  Flux density  \\
          & (\AA)   &  ($\times10^{-16}\ergpscmpspa$)  \\
\hline
$UVW2$ & 2025 & $12.6\pm0.5$ \\
$UVM2$ & 2250 & $1.62\pm0.22$ \\
$UVW1$ & 2825 & $1.19\pm0.09$ \\
$U$    & 3500 & $1.36\pm0.06$ \\
$B$    & 4450 & $1.44\pm0.06$ \\
\hline
\label{tab:uv}
\end{tabular}
\end{center}
\end{table}

The OM operated in imaging mode, collecting data in $U$ ($3000-3900\A$), 
$B$ ($3900-4900\A$), $UVW1$ ($2450-3200\A$), $UVM2$ ($2050-2450\A$) and
$UVW2$ ($1800-2250\A$).  Measured count rates in each filter were converted
to flux densities following Chen (2004).  Fluxes are given for each OM
filter in Table~\ref{tab:uv}.

\subsection{ATCA radio observations}
\label{sect:atca}
The source was observed with the Australia Telescope Compact Array (ATCA) 
at 1.4, 2.4, 4.8, 4.9, 5.1, 8.6, 17.7, 17.9 GHz on 2004 April 12--13 in the 
'snapshot' mode, for a 
total of 4 hours. The observations were performed with 
a total bandwidth of 128 MHz using the EW367 array configuration. The 
source was observed using 8 different frequency settings,
each time simultaneously at 2 frequencies.

The detected visibilities were calibrated, flagged and imaged using the 
MIRIAD data reduction package.
Flux density and bandpass calibration were carried out using 
PKS~B1934--638, and the same source was also used to correct for changes in 
gain and phase.
\begin{table}
\begin{center}
\caption{ATCA radio observations of \pks.  Frequencies ($\nu$) and corresponding
flux densities ($f_\nu$) are given in columns 1 and 2, respectively.
}
\begin{tabular}{cc}
\hline
(1) & (2)  \\
Observed &    Flux density  \\
frequency (GHz)       &  (mJy)  \\
\hline
1.380 & $791\pm28$ \\
2.372 & $705\pm12$ \\
4.796 & $486\pm34$ \\
4.924 & $469\pm15$ \\
5.052 & $446\pm15$ \\
8.636 & $333\pm13$ \\
17.732 & $230\pm10$ \\
17.861 & $225\pm9$ \\
\hline
\label{tab:rad}
\end{tabular}
\end{center}
\end{table}

We used MIRIAD task {\sc invert} to produce images, using uniform weighting to 
suppress side-lobes but without significantly degrading the beam size 
(sup=1, robust=0.5 in {\sc invert}). 
The produced dirty maps were then CLEANed and 
restored. The MIRIAD task {\sc imfit} was used to determine the integrated flux 
densities at each frequency, and the results are shown in Table~\ref{tab:rad}.

\subsection{Ceduna radio monitoring campaign}
\label{sect:ced}
The University of Tasmania $30\m$ Ceduna radio telescope was used to
intensively monitor the flux density of the source at 6.65~GHz over
the period 2004 February 5 - July 20.  During this period \pks\
was included in the list of sources observed as part of the COSMIC
project to monitor intraday variability (McCulloch \et 2005).  The
observations were made with a receiver with two orthogonal circular
polarizations operating in the frequency range $6.4-6.9$~GHz with a
system equivalent flux density of approximately 550~Jy.  The COSMIC
project alternates (with a typical period of $10-14$ days) between
monitoring two different groups of sources, one to the north of the
zenith at Ceduna, the other to the south.  \pks\ was included
in the southern source list and was observed 30 or more times every
day at regular intervals when above an elevation of $10^\circ$.  The
amplitude scale of the observations were set through regular
observations of the calibrator PKS~B1934--638 which was assumed to have
a flux density of 3.92 Jy at 6.65~GHz.  The RMS measurement error in a
single flux density measurement for COSMIC observations of a source
with a mean flux density of 400 mJy is approximately 35 mJy (McCulloch \et
2005).

\subsection{Siding Spring optical and NIR observations}
\label{sect:ss}
\pks\ was observed with the Australian National University (ANU)
1-metre telescope at Siding Spring Observatory on the nights of April
10--12 inclusive. The source rose in the middle of the night, and was
followed from $\approx$14:30~UT until twilight -- a period of
approximately five hours each night.

Observations were all $300\s$ in length, and were reduced using
standard techniques (bias-subtraction, bad-pixel removal and
flat-field correction).  Photometry of the source was measured in
$BVRI$ bands, and referenced to the standard stars of Graham (1982)
(using those in the E8 region, only $\sim20$~arcsec away from the
source). The photometry is listed in Table~\ref{tab:ir}. There is some
night-to-night variation, particularly in $B$ and $I$ bands, although
comparison with field stars suggests at least some of this variation
may be due to changes in the atmospheric transmission rather than
intrinsic to the source.

Observations were also made in the near-infrared at $JHK$ bands, using
the {\it CASPIR} instrument (McGregor \et 1994) on the ANU $2.3\m$
telescope, on the night of the multiwavelength observations (April
11). Images were made from four multiple dithered 60s exposures, each
made up of 2$\times$30s exposures in $J$, 6$\times$10s exposures in
$H$, and 12$\times$5s exposures in $K$, and were again reduced with
standard techniques (e.g. see Whiting \et 2002). 

As well as obtaining the photometry on each night, we tracked the
source with sequential $R$-band images (of 300~sec exposure), to
search for the presence of rapid variability. Relative photometry was
performed using several nearby field stars, but no significant
variations of \pks\ were seen at the precision attained.

\begin{table}
\begin{center}
\caption{Siding Spring optical and near infrared observations of \pks.  
Filters and corresponding wavelengths 
are given in columns 1 and 2, respectively.  Flux densities 
in each filter for nights 1, 2, and 3, are reported in column 3, 4, and
5, respectively.  Night 2 (April 11) is quasi-simultaneous with the \xmm\ and
ATCA observations.  
}
\begin{tabular}{ccccc}
\hline
(1) & (2) & (3) & (4) & (5) \\
Filter & $\lambda$ &  Night 1 & Night 2 & Night 3  \\
       & (\AA)     &  \multicolumn{3}{|c|}{($\times10^{-16}\ergpscmpspa$)}  \\
\hline
$B$    & 4400  & 1.13 $\pm$ 0.51  & 1.29 $\pm$ 0.28  & 2.09 $\pm$ 0.22  \\
$V$    & 5500  & 1.61 $\pm$ 0.19  & 1.63 $\pm$ 0.16  & 1.67 $\pm$ 0.11  \\
$R$    & 7000  & 1.38 $\pm$ 0.27  & 1.50 $\pm$ 0.17  & 1.67 $\pm$ 0.20  \\
$I$    & 8800  & 0.72 $\pm$ 0.10  & 0.78 $\pm$ 0.09  & 0.87 $\pm$ 0.13  \\
$J$    & 12390  & ---              & 1.24 $\pm$ 0.23  & ---\\
$H$    & 16490  & ---              & 0.77 $\pm$ 0.13  & ---\\
$K$    & 21320  & ---              & 1.08 $\pm$ 0.04  & ---\\
\hline
\label{tab:ir}
\end{tabular}
\end{center}
\end{table}

\subsection{Mt Canopus optical observations}
\label{sect:mtc}

The optical counterpart of \pks\ was observed with the Mt Canopus 1-m
telescope on 3 nights, 2004 April 3, 11 and 12. Measurements were made
using an SITe backside thinned 512x512 pixel CCD with Cousins $B$
filter. Magnitudes were determined relative to 3 stable reference stars in
the CCD field. In the absence of any measurements of standard stars we
applied an offset to equate the magnitude measured on 2005 April 11 to the
calibrated magnitude measured at the AAT on the same night
(see Table~\ref{tab:can}).

The source increased in brightness by 0.2 magnitudes between the first and
second measurements and by a further 0.15 magnitudes on the third
measurement. These changes are highly significant statistically. A similar
trend is also apparent in the AAT measurements from April 11-12.
\begin{table}
\begin{center}
\caption{Mt Canopus optical $B$ filter observations of \pks\ on
2004 April 3, 11 and 12.  In the absence of any measurements of standard stars, 
an offset was applied to equate the magnitude measured on 2005 April 11 to the
calibrated magnitude measured at the AAT on the same night. 
Date and magnitude is given in column 1 and 2, respectively.
}
\begin{tabular}{cc}
\hline
(1) & (2)  \\
Date & $B$  \\
\hline
April 3 & $19.42\pm0.03$  \\
April 11 & $19.22\pm0.04$  \\
April 12 & $19.07\pm0.03$  \\
\hline
\label{tab:can}
\end{tabular}
\end{center}
\end{table}

\section{Radio analysis}
\label{sect:ra}

\subsection{Spectral analysis}
\label{sect:rsa}
ATCA pointings were obtained at 1.4, 2.4, 4.8, 4.9, 5.1, 8.6, 17.7 and
17.9 GHz.  A power law fit to the spectrum produced a spectral slope of
$\alpha=0.52$.  Flux densities are given in Table~\ref{tab:rad}. 
Calculating the radio loudness ($R=f_{4.9GHz}/f_{B}$, Kellerman \et 1989) from 
simultaneous optical and radio data resulted in $R=3800$ confirming the 
high radio-loudness of this object.

\subsection{Radio variability}
\label{sect:idv}

\begin{figure}
\rotatebox{-90}
{\scalebox{0.32}{\includegraphics{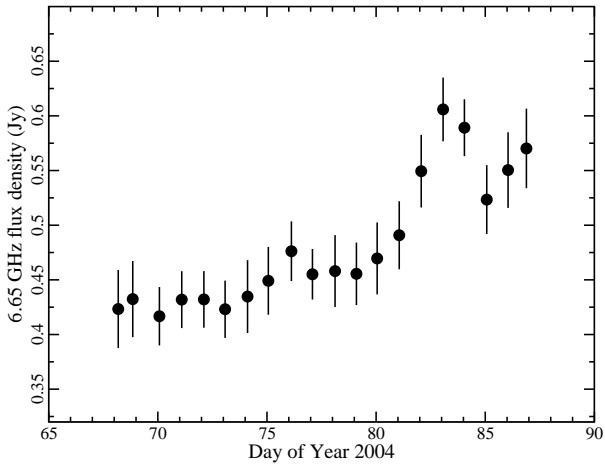}}}
\caption{A segment of the $6.65$~GHz light curve from the Ceduna monitoring
campaign of \pks, which lasted nearly six months.  
The section is of the most variable interval
between days $68-87$.
}
\label{fig:cedlc}
\end{figure}

For the majority of the nearly six month period (2004 February 5 - July 20)
that Ceduna monitored \pks\ the
source exhibited relatively little variability; however, in the period
just prior to the \xmm\ monitoring significant variations were
observed at radio wavelengths.  Over a period of 19 days between 2004
March 10-29 (day 68--87 of year 2004) the flux density at 6.65~GHz
increased by approximately one-third, from $420\pm30$~mJy to
$570\pm30$~mJy (Figure~\ref{fig:cedlc}).  If we assume that the observed variation
is intrinsic to the source (rather than due to a propagation effect
such as interstellar scintillation), then causality limits the linear
size of the varying component to be $c \times t_{obs}(1 + z)$.  So for
timescale of 19 days in a source at a redshift of 0.24 the linear size
of the variable component in the source frame is less than $6.1 \times
10^{11}$~km (0.02~pc).  We emphasis that the size of the entire radio
emitting region is likely on pc scales.  With our adopted cosmology, 
the luminosity distance for
\pks\ is $1.2 \times 10^{9}$~pc which requires the angular size
of the variable component to be less than 3.4~$\mu$as
(micro-arcseconds).  This implies that the brightness temperature of
the variable component in the source frame is $1.9\pm0.5 \times
10^{14}$K, well in excess of the inverse Compton limit of $10^{12}$K
(Kellermann \& Pauliny-Toth 1969).  There are two possibilities that
can account for this, one is that the jet has a Doppler boosting
factor of approximately 6 which reduces observed brightness
temperature to around the inverse Compton limit.  Alternatively, since
the Ceduna and ATCA observations show that approximately 10 days later
the radio flux density of the source had returned to its long-term
mean value of around 450~mJy, it may be that the component dissipated
over that period due to inverse Compton scattering.

The argument above assumes that the observed radio variations are due
to intrinsic changes in the source.  However, an angular size of the
order of a few microarcseconds for the variable component is
sufficiently small that we would expect it to show interstellar
scintillation (e.g. Bignall \et 2003).  In general the Ceduna light
curve over the period in question shows a steady increase in the flux
density of the source, however, there is a sudden change around day 81
of year 2004.  If intrinsic, this change of around 100~mJy on a
timescale of about 3 days implies brightness temperatures of around
$10^{15}$K, alternatively, it may be that this variation is in fact
due to scintillation of the component responsible for the 19-day
timescale change.  

Although we monitored \pks\ for
approximately half a year, significant variability was only detected
over a short period.  
CSS/GPS sources are typically not Doppler boosted, but there are 
examples to the contrary (e.g. O'Dea 1998; Aller \et 2002; Stanghellini \et 2005).
To our knowledge, ours was one of the most intense monitoring
campaigns of a CSS source, so it is difficult to determine if the
behaviour exhibited by \pks\ is characteristic of CSS sources.
In any regard, our campaign shows that daily radio frequency monitoring
of such sources from programmes such as COSMIC (McCulloch \et 2005)
can reveal phenomena that are likely to be missed by more traditional
monitoring where the interval between observations is weeks or months.

\section{X-ray spectral and timing analysis}
\label{sect:xst}

\subsection{Spectral analysis}
\label{sect:xsp}
We began the X-ray spectral analysis by fitting the 
observed $2-10\keV$ spectrum of each EPIC instrument
with a single power law.  Finding relatively good agreement
within the uncertainties we proceeded by fitting the three
spectra together.  The power law fit to the $2-10\keV$
EPIC spectrum resulted in a good fit ($\chidof=0.84/306$)
with a photon index of $\Gamma=1.44\pm0.06$.

\begin{figure}
\rotatebox{270}
{\scalebox{0.32}{\includegraphics{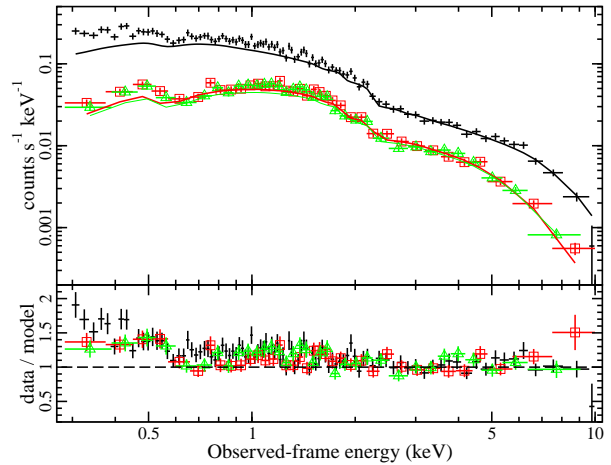}}}
\caption{A common power law ($\Gamma=1.44\pm0.06$) fitted
to the observed $2-10\keV$ pn (black crosses), MOS1 (red squares)
and MOS2 (green triangles) spectra.  The residuals from
extrapolating this fit to $0.3\keV$ are shown in the lower panel,
and indicate possible soft excess emission.  However, a single 
power law fit to the entire spectrum is statistically satisfactory.
}
\label{fig:po}
\end{figure}

Extrapolating this power law to $0.3\keV$ revealed an 
excess forming toward lower energies (Figure~\ref{fig:po}).
Even though refitting a single power law to the $0.3-10\keV$
spectrum ($\Gamma=1.56\pm0.02$) resulted in an acceptable
fit ($\chidof=0.97/792$) positive residuals remained below
$\sim0.6\keV$.

Replacing the single power law with a broken power law provided
significant improvement ($\Delta\chi^2=35$ for 2 additional free 
parameters) with an F-test probability of $\sim10^{-8}$.
In this case the high-energy and low-energy power laws had
$\Gamma=1.52\pm0.02$ and $\Gamma=2.10^{+0.23}_{-0.19}$, respectively,
with the break occurring at $E_b=613^{+79}_{-59}\eV$.

Additional neutral absorption at the redshift of \pks\ was considered
in all of the attempted models, but not found to be required.  
The $90$ per-cent upper limit on the column density of any such absorber
is $<1.6\times10^{19}\pscm$.
The existence of an emission feature at Fe\ka\ energies, which could
arise from a putative torus or accretion disc was also examined.
Again, considering such a feature did not significantly improve 
existing fits.  The $90$ per-cent upper limit on the equivalent width
of a narrow emission line between $6.4-7.0\keV$ was $EW<54\eV$.

The X-ray spectrum of \pks\ is unusually flat for a NLS1, but not
unlike radio-loud AGN.  There is a slight indication of a 
weak soft excess below $\sim1\keV$, which is typical of NLS1, but
also comparable to the broken power law seen in Flat-Spectrum Radio 
Quasar (FSRQ) and Low-energy-peaked BL Lacs.
On the other hand,
X-ray spectral studies of samples of GPS sources (Guainazzi \et 2006 and 
Vink \et 2006) do not show evidence of soft excess emission.
We note that statistically a single power law provides a satisfactory fit,
and that higher quality data are required to
pin down the existence of soft excess emission in \pks. 

From the broken power law fit to the EPIC data, an unabsorbed
$0.3-10\keV$ flux of $\sim1.5\times10^{-12}\ergpscmps$ is estimated.
The intrinsic $0.3-10\keV$ and $2-10\keV$ luminosities are
$2.42\times10^{44}$ and $1.54\times10^{44}\ergps$, respectively.

\subsection{X-ray variability}
\label{sect:xv}

\subsubsection{Short-term X-ray variability}
\begin{figure}
\rotatebox{270}
{\scalebox{0.32}{\includegraphics{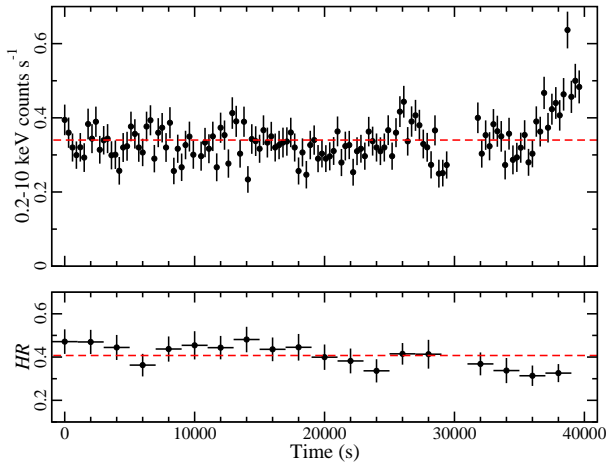}}}
\caption{Top panel:  The EPIC pn $0.2-10\keV$ light curve ($300\s$ bins). 
Lower panel: The hardness ratio ($HR=H-S/H+S$, where $H=0.7-10\keV$ and $S=0.2-0.7\keV$) variability
curve ($2000\s$ bins).  The energy bands were selected to examine the possible different
components predicted by the broken power law model (Section~\ref{sect:xsp}).
In both figures, the dashed lines mark the
average values.  Variability is negligible in the light curve with the
exception of the final $4\ks$ when a $30$ per cent flux increases is seen.
Though not highly significant ($\chi^2_{\nu}=1.05$), the $HR$ shows
softening over the course of the observation. 
}
\label{fig:lc}
\end{figure}

Short-term flux variations were examined over the $40\ks$ \xmm\
observation of \pks. The $0.2-10\keV$ light curve (Figure~\ref{fig:lc},
upper panel) shows some variability ($\chidof=1.95/124$), most of it 
appearing as a $30$ per cent
rise in count rate during the final $\sim4\ks$.

Spectral variations with time were searched for by constructing
a hardness ratio ($HR=H-S/H+S$, where $H=0.7-10\keV$ and $S=0.2-0.7\keV$) variability curve 
(Figure~\ref{fig:lc}, lower panel).
The energy bands were selected to examine the potentially components
on either side of the break energy predicted by the broken power law model
(Section~\ref{sect:xsp}).  The hardness ratio curve is not highly variable when
compared to a constant fit ($\chi^2_{\nu}=1.05$), but there is a softening 
trend over the course of the observation.  Therefore, it does seem plausible that
multiple components may be present in the X-ray continuum.

\subsubsection{Long-term X-ray variability}
The only other X-ray observation of \pks\ was a detection during
the \rosat\ All-Sky Survey (RASS; Voges \et 1999).
At that time it was detected as a faint source with a $0.1-2.4\keV$
count rate of $2.2\pm1.1\times10^{-2}\cps$ (Siebert \et 1998).  
Assuming a photon index in this energy
band of either 2.10 or 1.56 (as derived from our broken power law and
power law fits in Section~\ref{sect:xsp}), the unabsorbed $0.3-2.4\keV$
flux is $2.0\pm1.0$ or $2.5\pm1.2\times10^{-13}\ergpscmps$, respectively.
In this same energy band, the \xmm\ flux is 
$6.09^{+0.89}_{-1.06}\times10^{-13}\ergpscmps$.
This indicates that, at the very least, modest X-ray variability on about
the $30$ per cent level occurs on yearly time scales.
This level of variability is comparable with the centimetre fluctuations
on similar time scales seen in \pks.

\section{Contemporaneous spectral energy distribution}
\label{sect:sed}
\begin{figure}
\rotatebox{270}
{\scalebox{0.32}{\includegraphics{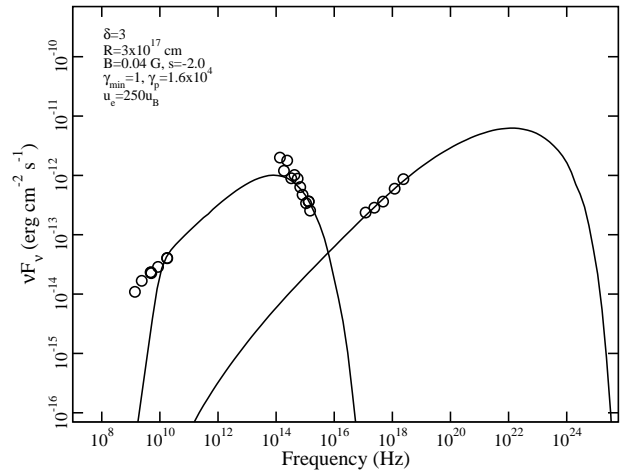}}}
\caption{Synchrotron/SSC scenario applied to the SED of \pks.
Model parameters are given in the top left.
For simplicity the X-ray data are shown as
five points between $0.3-10\keV$.  See text for details.
}
\label{fig:sed4}
\end{figure}

We considered the possibility that the SED of \pks\ is generated by 
synchrotron plus synchrotron self-Compton (SSC) processes.  Such
models are reasonably successful in describing the SED of blazars and
even some un-beamed radio galaxies (e.g. Chiaberge \et 2001).
The adopted model (Kataoka \et 1999) assumes a single zone homogeneous
SSC in a spherical volume, which is probably an oversimplification,
but allows the source to be modelled in a self-consistent way 
with a minimum of parameters, and is not
unreasonable given the compactness of \pks.
Given the modest radio variability 
(Sect~\ref{sect:idv}), models with large Doppler beaming factors are not
considered.  The model shown in Figure~\ref{fig:sed4}
reproduces the general trend of the SED quite well;  however it
significantly under estimates the radio emission.  
Such discrepancies in the radio are found in many studies and 
it is believed to be a consequence
of the radio emission originating from a much larger region than the
X-rays and gamma-rays.  Even so, the current region size ($R\approx0.1\pc$)
is too large to account for the 
short-term X-ray variability (Sect.~\ref{sect:xv}).  
Furthermore, explaining the high X-ray brightness would require a large
electron-to-magnetic energy density ratio ($u_{e}/u_{B}$) for the SSC
component.  In extreme objects such as blazars, the ratio is
typically less than 100, whereas for all the X-ray emission to be produced
by the single zone SSC model in \pks\
the ratio would have to be about 250.

We have examined whether the SED can be constrained by data at higher
energies. Mattox \et (2001) made a quantitative evaluation of potential radio
identifications for 3EG EGRET sources based on the 5 GHz flux density,
spectral index, and proximity to the EGRET source of candidate
counterparts. \pks\ was considered as a counterpart to the
unidentified EGRET source 3EG J1958$-$4443, however the association had an
a posteriori probability of less than $2\times 10^{-4}$ and so was
considered unlikely (although no more likely candidates were found).
Even considering the SED model with a very high $u_{e}/u_{B}$, the
predicted $10^{23}\Hz$ flux density falls
below the detection sensitivity of EGRET. 

\begin{figure}
\rotatebox{270}
{\scalebox{0.32}{\includegraphics{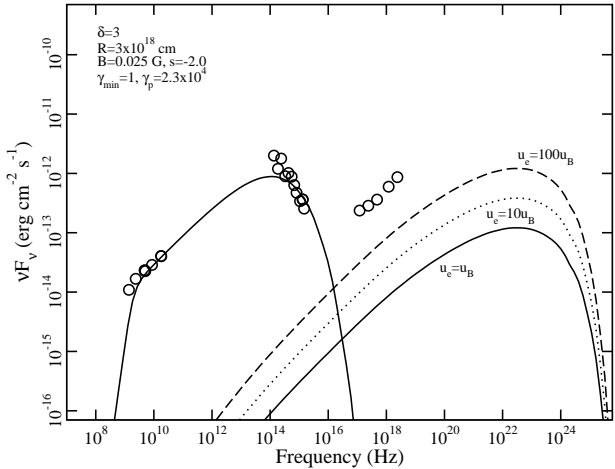}}}
\caption{Synchrotron/SSC scenario applied to the SED of \pks.
The emitting region is much larger than in Figure~\ref{fig:sed4},
which better describes the radio emission.  The SSC is described
with more typical values of $u_{e}/u_{B}$.  In this case, 
a component in addition to SSC is required to fit the high X-ray
brightness.
}
\label{fig:sed5}
\end{figure}

A synchrotron/SSC model, which adopts a much larger emitting region
characterises the radio-to-UV emission quite well (Figure~\ref{fig:sed5}).  
On the other hand, assuming equipartition conditions or moderate
particle-dominated parameters, the SSC model under estimates the observed
X-ray emission by an order of magnitude (Figure~\ref{fig:sed5}).
This may not be unreasonable if the SED is a combination of jet emission
and disc/halo emission as typically seen in Seyfert 1s.  Indeed the 
short term X-ray variability is consistent with originating from a much
smaller region than most of the radio emission.

\begin{figure}
\rotatebox{270}
{\scalebox{0.32}{\includegraphics{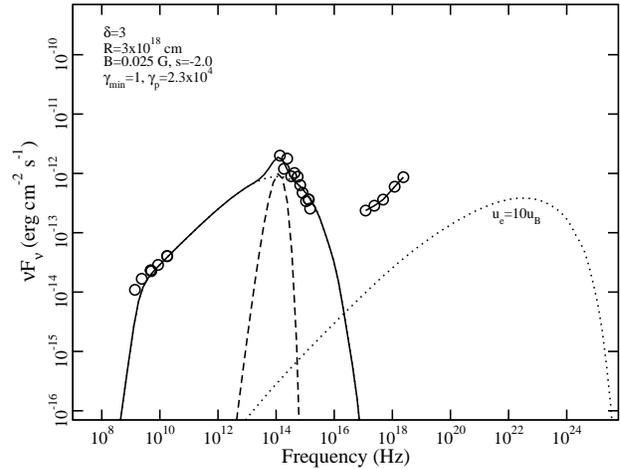}}}
\caption{A possible interpretation for the SED of \pks\ (black solid line
is the sum of all components).
Most of the underlying continuum is produced by the synchrotron/SSC 
mechanism (dotted line).  The IR excess is attributed to $1000-1500\K$ blackbody
emission from the AGN heated torus (dashed line).  The strong X-ray excess 
is generated by thermal Comptonisation processes close to the central
black hole.
}
\label{fig:sed6}
\end{figure}

We probe this further by re-examining the \xmm\ X-ray data.
The data are refitted with a thermal Comptonisation model (Titarchuk 1994)
plus an underlying power law to mimic the SSC component.  The 
photon index of the SSC power law is fixed to $\Gamma=1.52$ (similar to
the slope of the radio spectrum), while the normalisation is allowed to
vary freely.  This produces a good fit ($\chidof=0.97/790$) similar to the
broken power law model in Section~\ref{sect:xsp}.  The model parameters:
plasma temperature ($T_e\approx54\keV$), optical depth ($\tau\approx1.12$)
and seed photon temperature ($kT=66\pm17\eV$) are not inconsistent
with physical conditions expected in AGN where the seed photons 
provided by the accretion disc are up-scattered to higher energies by 
the hot electron corona.  Of course, the values are uncertain given the
known degeneracy between $T_e$ and $\tau$.
In this interpretation the flux density of the underlying power law is 
$1.2\pm0.4\times10^{-14}\ergpscmpspkeV$ at $2\keV$, consistent with originating
from the SSC component with $u_{e}\approx10u_{B}$ (Figure~\ref{fig:sed6}).

A second issue is the slight excess seen in the infrared data
compared to the synchrotron models shown in Figures~\ref{fig:sed4} 
and~\ref{fig:sed5}.  de Vries \et (1998) recognised this IR excess in other
CSS sources and attributed it to a $\sim1000\K$ blackbody arising from
an AGN heated torus.  We draw a similar conclusion for the IR emission
in \pks, which can be fitted by a blackbody with a temperature of
about $1000-1500\K$ (Figure~\ref{fig:sed6}).  

Overall, the inclusion of these 
additional component to the synchrotron/SSC continuum provides a
reasonable approximation of the SED of \pks\ (solid line in
Figure~\ref{fig:sed6}).

\section{Discussion}
\label{sect:dis}

\subsection{The SED of \pks}
The shape of the SED of \pks\ portrays the typical double-hump form
that is often seen in radio-loud, blazar-type AGN.  The single zone homogeneous
synchrotron/SSC interpretation appears to be a good base model.  However, 
the fact that the normally constant, long-term radio light curve shows 
occasional activity and because the X-ray emission is apparently excessive,
indicate that a single zone model is probably an oversimplification. 

\subsubsection{General SED characteristics}
Utilising the simultaneous UV data we calculated the optical-to-X-ray
spectral slope ($\alpha_{ox}$) of \pks\ to be approximately $-0.95$.  
Considering
the UV luminosity dependence of $\alpha_{ox}$ for radio-quiet, unabsorbed
AGN (e.g. NLS1) a value of $\alpha_{ox}=-1.27$ is predicted (Strateva \et 2005)
indicating that the spectral slope of \pks\ is flat compared to NLS1-type 
objects.

We also calculated the spectral slopes between the radio and optical 
($\alpha_{ro}$), and radio and X-rays ($\alpha_{rx}$) as shown
by Padovani \et (2002) for a sample of blazar-type objects.
In comparing the values of all the spectral indices measured for \pks\
to the Padovani sample, it is found that the characteristics of \pks\
are very similar to FSRQ.
Essentially, \pks\ resembles a radio-loud quasar, but
with several atypical properties (e.g. IR excess but no intrinsic X-ray
absorption, normally low radio and X-ray variability, NLS1 classification),
and from the SED modelling the only way to recover a typical blazar SED
is with a very large $u_{e}/u_{B}$.

\subsubsection{The X-ray excess}
As seen in Figures~\ref{fig:sed4} and \ref{fig:sed5} simple synchrotron/SSC 
models are unable to simultaneously fit the radio and X-ray emission.
The shortcoming is apparently the difficulty in fitting the
``too bright'' X-ray emission in \pks.  
The high X-ray brightness could be explained
by a large $u_{e}/u_{B}\approx250$.  However, this is quite unusual
compared to other types of jet-dominated objects.

Alternatively, the X-ray emission could require an additional component, which
originates close to the black hole.  The presence of some short-term X-ray 
variability and normally
constant
radio variability implies that the X-ray emitting region is probably
much smaller than
the radio one. 
The SED of \pks\ is well approximated if in addition to a
synchrotron/SSC continuum, the model takes on a Comptonisation component to
describe the strong X-ray emission  
(empirically, a single or broken power law fit the X-rays adequately).
In this case, the derived model parameters are not atypical of expected physical
conditions in AGN.

\subsubsection{Excess optical reddening}
Excess emission in the IR has been seen in other CSS sources (de Vries \et 1998);
therefore its presence in \pks\ is not unexpected.  This excess can be
fit with a $1000-1500\K$ blackbody, which is consistent with 
originating in an AGN heated torus, but notably also close to the
sublimation temperature of dust.  The host galaxy cannot be excluded for
contributing to this component.  However, the optical spectra shown in Osh01
do not exhibit any striking galactic features, so we consider it unlikely that the
host galaxy dominates this spectral region.

The absence of big blue bump emission at UV energies could imply reddening of
the optical spectrum by this component. 
The Balmer decrement (\halpha/\hbeta) was measured in \pks\ using the
low-resolution optical spectrum shown in figure 2 of Osh01.  The ratio was
\halpha/\hbeta$\approx6$ for the narrow-line region (NLR) lines, with 
considerable uncertainty arising from the difficulty in isolating the
narrow component from the broad component in the fit ($\sim60$ per cent).
Assuming that the narrow Balmer
lines are due to case B recombination then by following the example of
Veilleux \et (1999), and using the parameterised reddening curve for our Galaxy 
(Miller \& Mathews 1972) the amount of intrinsic extinction in the NLR of \pks\
can be estimated.  If the ratio of total to selective extinction is 3.1
then the visible extinction in \pks\ is $A_V=1.9\pm1.5$.
Even within the large range of uncertainty the extinction is significantly 
higher than the estimated intrinsic column density found
in the X-ray spectral fits ($\nh<1.6\times10^{19}\pscm$).  Following Predehl \&
Schmitt (1995), the X-ray measurements give $A_V<0.01$.

The discrepancy could be understood if \pks\ had a very different 
gas-to-dust ratio than our Galaxy. 
This could perhaps be achieved as the
radio jet, which likely terminates on NLR-scales in \pks, deposits material
from the nucleus into the NLR.  It would be of interest to examine this issue 
in other
CSS/GPS sources to determine if the discrepancy in X-ray and optically measured
$A_V$ is a class property.

\subsection{\pks: NLS1, CSS, or both?}

The radio properties of \pks\ satisfy the CSS definition of the source.
However, the optical spectral properties are not entirely consistent
with the general behaviour of CSS sources (e.g. Gelderman \& Whittle 1994,
Morganti \et 1997).  Osh01 proposed a NLS1 classification for \pks\
because of its relatively weak \oiii\ emission and narrow \hbeta\ line.
The classification has prompted some debate primarily due to the
lack of strong \feii\ emission in \pks, and it
has been suggested that \pks\ is probably a type 2 AGN (e.g. Sulentic \et 
2003, Zhou \et 2003).
However, the X-ray emission is unabsorbed, and there is evidence for
short-term variability and a weak soft excess.  Moreover, the \oiii\ emission
is relatively weak and the \hbeta\ emission is variable on at least
yearly time scales (Osh01) indicating that the broad-line region is probably
visible.  For the sake of argument, in this discussion we will treat \pks\
as a NLS1.

As such, there are four rather securely identified radio-loud NLS1
and perhaps two more questionable objects (\pks\ and RGB~J0044+193),
which have been relatively well examined.  Nonetheless, of
particular interest is that five of these six objects display steep or
inverted radio spectra, which are consistent with radio definitions of 
CSS/GPS sources.
This coincidence in definitions presents a challenging area of research,
for a number of reasons.  First, radio-loud AGN and NLS1 occupy distinctly
different regions of the primary eigenvector 1 parameter space
(Boroson \& Green).  NLS1 are typically very radio-quiet.  Secondly,
the fraction of radio-loud NLS1 which are CSS/GPS sources (5/6) is potentially
much higher than the fraction of CSS/GPS sources in radio selected sample
($10-30$ per cent).  Thirdly, both classes of objects are believed
to be in the early stages of AGN evolution.
This implies that radio-loud NLS1 could be CSS/GPS sources, perhaps in a 
situation where the 
black hole environment and the radio component are forming simultaneously.

\begin{figure}
\rotatebox{270}
{\scalebox{0.32}{\includegraphics{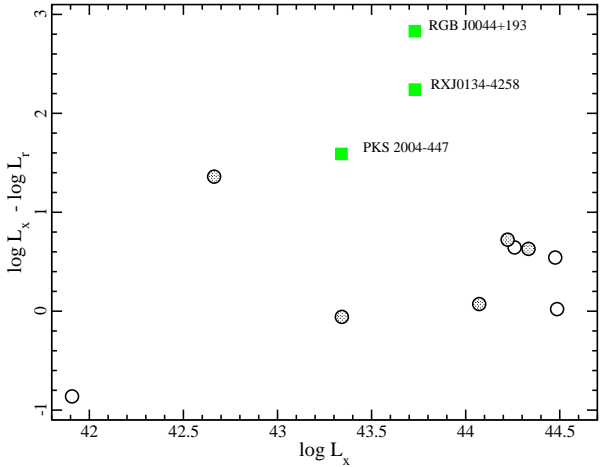}}}
\caption{The $2-10\keV$ and $5$~GHz luminosity
ratio plotted against $2-10\keV$ luminosity ($\ergps$).
The circles represent the GPS samples of Guainazzi \et (2006; empty circles)
and Vink \et (2006; shaded circles).  The green, filled squares represent
the radio-loud NLS1-CSS/GPS objects, which are also designated as GPS/CSS sources.  
There is possible indication that the NLS1-CSS/GPS are not X-ray weak like
typical GPS sources.
The luminosity of \pks\ appears different in
the figure than in the text because the cosmology was changed to match the
luminosities of the larger surveys. 
}
\label{fig:lxlr}
\end{figure}

\begin{figure}
\rotatebox{270}
{\scalebox{0.32}{\includegraphics{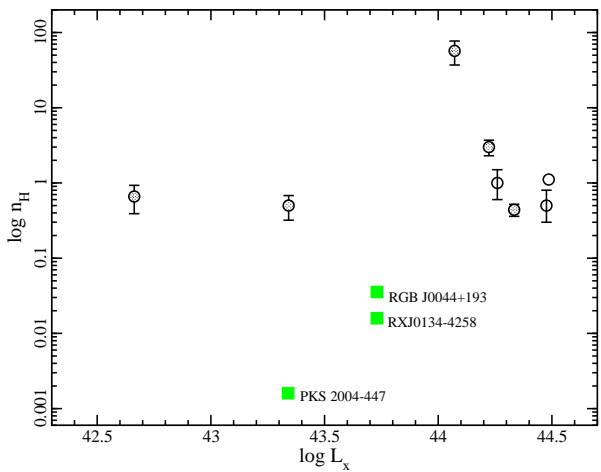}}}
\caption{Intrinsic neutral hydrogen column density as a function of
$2-10\keV$ luminosity ($\ergps$).  The symbols are as in Figure~\ref{fig:lxlr}.
Symbols without error bars in the column density indicate upper limits.
There is some indication that the NLS1-CSS/GPS objects possess less absorption
at a given X-ray luminosity than the GPS sources.  
The luminosity of \pks\ appears different in
the figure than in the text because the cosmology was changed to match the
luminosities of the larger surveys.
}
\label{fig:nhlx}
\end{figure}

We compared some of the NLS1-CSS/GPS sources mentioned above with 
X-ray samples of GPS objects (from Vink \et 2006 and Guainazzi \et 2006)
to determine how similar the two groups appear.
Our findings are illustrated in Figure~\ref{fig:lxlr} and \ref{fig:nhlx}
where the NLS1-CSS/GPS objects (\pks, RXJ0134--4258, RGB~J0044+193) are shown
as filled squares (a common cosmology is adopted for the figures).  

In Figure~\ref{fig:lxlr} we note that the NLS1-CSS/GPS objects appear
to have stronger X-ray emission compared to the radio emission than
typical GPS sources, which are known to be X-ray weak.
This is in line with the additional X-ray component required to fit the
SED of \pks.  The significant, rapid X-ray variability seen in the
three objects (see Grupe \et for RXJ0134--4258 and Siebert \et for RGB~J0044+193)
indicate that, like \pks, the X-ray emission region is not extended.

In Figure~\ref{fig:nhlx} a comparison is made between the amount of intrinsic
absorption seen in NLS1-CSS/GPS and the GPS sample.  There is support that
lower intrinsic absorption is seen in the NLS1-CSS/GPS, independent of 
X-ray luminosity.

Significant statements cannot be made given the small samples presented
in Figure~\ref{fig:lxlr} and \ref{fig:nhlx}; however, there is an indication
that NLS1-CSS/GPS and `typical' GPS sources differ in X-ray behaviour.

\section{Summary}
\label{sect:conc}

An investigation of the SED of \pks, a CSS and possible radio-loud NLS1, is
presented, utilising contemporaneous X-ray, near-UV, optical, near-IR, and
radio data.  The main results are summarised here.

\begin{itemize}
\item
The SED cannot be fitted by a single zone homogeneous SSC model.  Primarily,
the difficulty is in simultaneously fitting the radio and excessive X-ray 
emission.
Models in which a second component (probably arising from Comptonisation 
of accretion disc photons) is adopted to describe the X-ray emission or
SSC models that require very high values of $u_{e}/u_{B}$ seem 
plausible.

\item
Excess emission is also seen in the SED at IR energies.  In fact, the peak
emission in the SED appears to be in the IR.  This excess can be well
fitted with a $T\approx1000\K$ blackbody, likely originating from an AGN
heated torus; however we cannot exclude some contribution from the host 
galaxy. 
No excess absorption (above Galactic column density)
or fluorescent iron line is detected in the X-ray spectrum, as expected if
the X-rays transverse a high column density.
Analysis of a non-contemporaneous, low-resolution optical spectrum suggests
that the NLR is much more reddened than the X-ray emitting region.
This could arise if the gas-to-dust ratio in \pks\ is very different than in
our own Galaxy, but further analysis of higher quality optical spectra are
required to draw firm conclusions.

\item
Long-term radio monitoring of \pks\ shows a rather constant light curve
over a six month period with the exception of one event when
the $6.65$~GHz flux increases by $\sim35$ per cent in 19 days.  
Interstellar scintillation cannot be excluded as the cause of the
variability, but the detection of the rare event demonstrates the 
potential of intense monitoring campaigns such as COSMIC
(McCulloch \et 2005).
No matter whether the outburst has an intrinsic or extrinsic origin,
it indicates a small emitting region for at least some of the radio
flux.

\item
There is some indication of a possible connection between NLS1 and CSS/GPS
sources.  In comparison to other GPS sources, these radio-loud NLS1 seem to
exhibit greater X-ray emission relative to radio flux,
and lower intrinsic absorption.
However, a larger and more uniformly selected sample is required to draw 
convincing conclusions.

\end{itemize}

The recent discovery of a large number of radio-loud NLS1 (Whalen \et 2006,
Komossa \et 2006b)
holds many possibilities for future investigations of the potential 
connection between radio-loud NLS1 and CSS/GPS sources.


\section*{Acknowledgements}

Based on observations obtained with \xmm, an ESA science mission with
instruments and contributions directly funded by ESA Member States and
the USA (NASA). 
The Australia Telescope Compact Array is part of the Australia Telescope
which is funded by the Commonwealth of Australia for operation as a
National Facility managed by CSIRO.
LCG acknowledges funding from the Japan Society for the Promotion of
Science through a JSPS Postdoctoral Fellowship, and would like to thank
Patricia Arevalo for helpful discussions.
SPE would like to thank the Australian Research Council for financial
support for this work, Steve Carter and Cliff Senkbeil for their
assistance collecting and processing the data, and Jim Lovell and Giuseppe
Cim\`o for useful discussions about the nature of the Ceduna variability.
Thanks to Ivy Wong for assistance with the optical observing, Steve
Longmore for taking the near-infrared CASPIR observations, and the anonymous
referee for a quick and helpful report.

\bsp
\label{lastpage}
\end{document}